\documentclass[11pt]{article}
\newcommand{\Springer}[1]{}
\newcommand{\Preprint}[1]{#1}

\usepackage{url}
\usepackage{amsmath,amssymb,graphicx,color}
\usepackage{array}
\usepackage{dcolumn}

\newcommand{\Init}{I}
\newcommand{\Unsafe}{U}

\newcommand{\Reset}{\rho}
\newcommand{\Inv}{\mathrm{Inv}}
\newcommand{\Barrier}{\nabla}
\Springer{\newcommand{\RR}{\mathbb{R}}}
\Springer{\newcommand{\hs}{\hspace*{0.3cm}}}

\newcommand{\SampledConstr}[1]{C_{#1}}

\newcommand{\Comment}[1]{}

\Preprint{\usepackage {ifthen}

\ifthenelse{\isundefined{\definition}}{\newtheorem{definition}{Definition}}{}

\ifthenelse{\isundefined{\theorem}}{\newtheorem{theorem}{Theorem}}{}
\ifthenelse{\isundefined{\conjecture}}{\newtheorem{conjecture}{Conjecture}}{}
\ifthenelse{\isundefined{\lemma}}{\newtheorem{lemma}{Lemma}}{}
\ifthenelse{\isundefined{\corollary}}{\newtheorem{corollary}{Corollary}}{}
\ifthenelse{\isundefined{\example}}{\newtheorem{example}{Example}}{}
\ifthenelse{\isundefined{\property}}{\newtheorem{property}{Property}}{}
\ifthenelse{\isundefined{\proof}}{\newenvironment{proof}{\noindent{\bf Proof.}\hspace{0.05in}}{\par\mbox{}\par}}{}

\newcommand{\vs}{\vspace*{0.3cm}}
\newcommand{\hs}{\hspace*{0.3cm}}
\ifthenelse{\isundefined{\qed}}{\newcommand{\qed}{\mbox{$\blacksquare$}}}{}

\hyphenation{}

\setlength{\unitlength}{6pt}

\newcommand{\NN}{\mathbb{N}}

\newcommand{\RR}{\mathbb{R}}

\newenvironment{lists}[1]{
                 \begin{list}{}{
                     \setlength{\listparindent}{0in}
                     \settowidth{\labelwidth}{#1}
                     \setlength{\leftmargin}{\labelwidth}
                     \addtolength{\leftmargin}{\labelsep}
                     }
                 }{
                 \end{list}
                 }

\newenvironment{given-find}[2]{
                               \vs 
                               \noindent \hrule
                               \begin{lists}{Given:XX}
                               \item[\sc Given: \hfill] #1                                 
                               \item[\sc Find: \hfill] #2                               
                               \vs 
                               \noindent \hrule 
                               }{
                               \end{lists}
                               }

}

\title{Simulation Based Computation of\\Certificates for Safety of Hybrid Dynamical Systems}
\author{Stefan Ratschan, \Preprint{\\Institute of Computer Science, Czech Academy of Sciences}\\ORCID: 0000-0003-1710-1513}
\Springer{\institute{Institute of Computer Science, Czech Academy of Sciences}}

\begin{document}

\maketitle

\begin{abstract}
  In this paper, we present an algorithm for synthesizing certificates---so-called barrier certificates---for safety of hybrid dynamical systems. Unlike the usual approach of using constraint solvers to compute the certificate from the system dynamics, we synthesize the certificate from system simulations. This makes the algorithm applicable even in cases where the dynamics is either not explicitly available, or too complicated to be analyzed by constraint solvers, for example, due to the presence of transcendental function symbols. 

The algorithm itself allows the usage of heuristic techniques in which case it does not formally guarantee correctness of the result. However, in cases that do allow rigorous constraint solving, the computed barrier certificate can be rigorously verified, if desired. Hence, in such cases, our algorithm reduces the problem of finding a barrier certificate to the problem of formally verifying a given barrier certificate.

\end{abstract}

\section{Introduction}

A common technique in formal verification is the reduction of a verification problem to a constraint solving problem. A main limitation of such approaches comes from theoretical and practical limitations of the decision procedures used to solve the resulting constraints. In the case of continuous, or hybrid systems, this is usually the theory of the real numbers which is undecidable as soon as periodic function symbols, such as the sine function are allowed. Even in the polynomial case, which is decidable~\cite{Tarski:51}, existing decision procedures are by far not efficient enough to be able to solve realistic problems. In contrast to that, simulations of continuous systems, approximating the solutions of the underlying differential equations, are possible for systems far beyond those restrictions.

In this paper, we circumvent the constraint solving bottleneck by using an approach that is data-driven instead of deductive: We use simulation data instead of system dynamics as the main input for computing certificates. From a given set of simulations we compute a candidate for a certificate. If this candidate turns out to not to be a certificate for the system itself, we use a refinement loop to run further simulations. In our concrete case, the certificates are formed by so-called barrier certificates~\cite{Prajna:04}. Our algorithm handles hybrid dynamical systems with non-deterministic dynamics in the form for disturbance inputs. 

The algorithm uses numerical optimization as its main workhorse. Here, we allow sub-optimal results which enables the use of fast heuristic~\cite{Locatelli:13} and numerical~\cite{Nocedal:99} optimization algorithms. 
In cases, where the system dynamics can be handled by rigorous decision procedures, the final result can be rigorously verified. This final verification step is then applied to a  barrier certificate that is already given. Hence it is a much easier problem than the computation of the barrier certificate itself. 
In our experiments, the non-verified results always turned out to  be mathematically correct. Moreover, the final rigorous verification step always took negligible time. The experiments also show that the approach can compute barriers for ordinary differential equations of a complexity that has been out of reach for computation of barrier certificates up to now.

\Springer{
The structure of the paper is as follows: In Section~\ref{sec:problem-description} we define the problem that the paper solves. In Section~\ref{sec:framework} we describe the general algorithmic framework. In Sections~\ref{sec:candidate} and~\ref{sec:cntrxpl} we show how to concretize this framework into a working algorithm, and in Section~\ref{sec:algorithm} we provide this algorithm. In Section~\ref{sec:implementation} we describe our implementation of the algorithm, in Section~\ref{sec:experiments} we describe computational experiments on several examples, and in Section~\ref{sec:related-work} we discuss related work. Section~\ref{sec:conclusion} concludes the paper.}



The research published in this paper was supported by GA{\v C}R grant GA15-14484S and by the long-term strategic development financing of the Institute of Computer Science (RVO:67985807). We thank Hui Kong for discovering a significant mistake in an earlier version of the paper.

\section{Problem Description}
\label{sec:problem-description}

\begin{definition}
  \label{def:2}
A \emph{(hybrid systems) safety verification problem} is a tuple $(M, \Omega, D, f, \Inv, \Reset, \Init, \Unsafe)$ where
\begin{itemize}
\item $M$ is a finite set (the \emph{modes} of the safety verification problem),
\item $\Omega\subseteq M\times\mathbb{R}^n$ (the \emph{state space} of the safety verification problem),
\item $D\subseteq \mathbb{R}^l$ where $l\in\NN_0$ (the set of \emph{disturbance inputs} of the safety verification problem), 
\item $f: (\Omega\times D)\rightarrow \mathbb{R}^n$, s.t. for every mode $m\in M$, the restriction of $f$ to $m$ is Lipschitz continuous (the \emph{dynamics}),
\item $\Inv\subseteq \Omega$ (the \emph{invariant})
\item $\Reset\subseteq \Omega\times\Omega$ (the \emph{reset relation}), 
\item $\Init\subseteq\Omega$ (the set of \emph{initial states}),  and
\item $\Unsafe\subseteq\Omega$ (the set of \emph{unsafe states}).
\end{itemize}
\end{definition}

This definition handles classical ordinary differential equations by allowing $M$ to contain only one dummy element, $f$ being independent of the disturbance input, $\Inv=\Omega$, and $\rho=\emptyset$. When clear from the context, we will use the term \emph{disturbance input} not only for the third item of Definition~\ref{def:2}, but for any Lipschitz continuous function $d: [0, T]\rightarrow D$.

Now we will give semantics to Definition~\ref{def:2}, first for continuous evolution:

\begin{definition}
  \label{def:4}
  Given a safety verification problem $H= (M, \Omega, D, f, \Inv, \Reset, I, U)$, for $x, x'\in \Omega$,
  disturbance input $d$, $T\geq 0$,   $x\xrightarrow{d,T}_H x'$ iff  $x'=y(T)$, where $y$ is a solution of the differential equation \[\forall t\in [0, T] \;.\; \dot{y}(t)= f(y(t), d(t)), y(t)\in \Inv\] with $y(0)=x$.
\end{definition}

In general, evolution can be continuous and discrete:

\begin{definition}
  \label{def:5}  
  Given a safety verification problem $H= (M, \Omega, D, f, \Inv, \Reset, I, U)$, for $x, x'\in \Omega$, $x\rightarrow_H x'$ iff either
there is a disturbance input $d$, $T\geq 0$ s.t. $x\xrightarrow{d,T}_H x'$, or  $(x, x')\in\Reset$.
\end{definition}

The condition $y(t)\in \Inv$ from Definition~\ref{def:4} restricts continuous evolution and allows us to enforce non-continuous behavior following the right-hand side of the disjunction in Definition~\ref{def:5}.

A flow from an initial to an unsafe state disproves safety:

\begin{definition}
    Given a safety verification problem $H= (M, \Omega, D, f, \Inv, \Reset, I, U)$, a \emph{counter-example} of $H$ is a sequence $x_1,\dots, x_n$ of elements of $\Omega$ s.t.
  \begin{itemize}
  \item $x_1\in\Init$,
  \item $x_1\rightarrow_H\dots \rightarrow_H x_n$, and
  \item $x_n\in\Unsafe$.
  \end{itemize}

\end{definition}

We want to verify that a given safety verification problem does not have a counter-example. The corresponding decision problem is in general undecidable~\cite{Bournez:08,Henzinger:98}, and decidable only for very special cases~\cite{Hainry:08}. Hence we head for an algorithm that successfully solves benchmark problems. 

The following object~\cite{Prajna:04,Taly:09} certifies successful safety verification: \Comment{\footnote{discuss difference to original definition}}
\begin{definition}
\label{def:1}
A \emph{barrier certificate} of a safety verification problem $(M, \Omega, D, f, \Reset, \Init, \Unsafe)$ is a  function $V: \Omega\rightarrow \RR$ such that for every $m\in M$ the restriction of $V$ to $m$ is continuously differentiable, and 
\begin{itemize}
\item $\forall x\in \Init \;.\; V(x)<0$, 
\item $\forall x\in \Unsafe \;.\; V(x)>0$, 
\item $\forall x\in\Inv, d\in D \;.\; V(x)=0 \Rightarrow (\nabla V(x))^T f(x, d)<0$,
where $\nabla V: \Omega\rightarrow \RR^n$ is s.t. for all $(m, x)\in\Omega$, $\nabla V(m, x)= \nabla V_m(x)$,   and 
\item $\forall\; (x,x')\in\Reset  \;.\; V(x)\leq 0 \Rightarrow V(x')< 0$.
\end{itemize}  
\end{definition}

In this paper, we will introduce an algorithm that, for an arbitrary given safety verification problem, tries to compute such a barrier certificate. If successful, this implies safety:

\begin{property}
  If a safety verification problem has a barrier certificate, then it has no counter-example.
\end{property}

Under the assumption of robustness and boundedness of the set of unsafe states, for differential equations without resets also the converse holds~\cite{Ratschan:18}.

Our approach is template based. That is, we introduce parameters into the function $V$, resulting in a 
parametric function $V(p, x)$ that we call \emph{template}. This reduces the problem of finding a barrier to the problem of finding parameter values such that the template is a barrier. The template can have an arbitrary form, but we will usually work with templates that are polynomial in each mode. This means for every mode $m\in M$,  it is of the form $p_{m,0}+\sum_{i=0}^{k_i} p_{m,i} \vec{x}^i$, where the $\vec{x}_i$ are power products, and the $p_{m,i}$ are parameters.


So, now we are left with the problem of finding a vector $p$ of parameter values such that
\begin{itemize}
\item $\forall x\in \Init \;.\; V(p, x)<0$,
\item $\forall x\in \Unsafe \;.\; V(p, x)>0$, 
\item $\forall x\in\Inv \;.\; V(p, x)=0 \Rightarrow (\nabla V(p, x))^T f(x)<0$, and
\item $\forall x, x'\in\Reset\;.\; V(p,x)\leq 0 \Rightarrow V(p, x')< 0$.
\end{itemize}

We denote the conjunction of these four constraints by $C_f$. The constraint $\exists p\; C_f$  represents a decision problem in the theory of real numbers with quantifier prefix $\exists\forall$. In the polynomial case, this is decidable~\cite{Tarski:51}, function symbols such as $\sin$ make the problem undecidable. However, even in the polynomial case, in practice, existing decision procedures can only solve problems with a few variables. 
Note also, that for a template with $k$ parameters, this constraint has $n+k$ variables.




In the rest of the paper we assume a fixed safety verification problem, $H=(M, \Omega, D, f, \Inv, \Reset, \Init, \Unsafe)$ and use accordingly simplified notation. 

\section{Algorithmic Framework}
\label{sec:framework}

Even if the dynamics $f$ is complex, it is usually possible to compute simulations of the system behavior by approximating the solution of the involved ordinary differential equations starting from a given initial value.
Simulation is an essential tool in practical systems modeling, and approximation is usually taken into account already during the modeling process. As a consequence, such simulations often describe the intended system behavior more accurately than even the precise mathematical solution.

For us, the important information will be the endpoints of such a simulation: We will call a pair $(s, s')\in\Omega\times\Omega$ s.t. there are $x_1,\dots,x_n\in\Omega$ with $s\approx x_1\rightarrow_H\dots\rightarrow_H x_n\approx s'$, a \emph{simulation segment}. When stating properties depending on such simulation segments below, we will assume that $\approx$ actually is equality. In practice, due to rounding and discretization errors, equality will not hold. However, we will design our method in such a way that such errors will not affect correctness of the method.


We will maintain a set $S$ of simulation segments. Our goal is to use this set $S$ for computing a solution $p$ of the constraint $C_f$. For this we relax the universal quantifiers to finite conjunctions. 
For the first two parts of the constraint $C_f$ we simply replace the set $I$ bounding the universal quantifiers in the first part with the set of all initial points in $S$, and the set $U$ with the set of all unsafe points in $S$. However, for the third part of $C_f$, due to the implication occurring here, it does not suffice to replace the set $\Omega$ by a finite subset. This would allow trivial satisfaction of this implication using a parameter vector $p$ such that $V(p, s)$ is non-zero for every element of this finite subset. Instead, we use the observation, that the third part of $C_f$---which ensures a certain direction of the vector field $f$ on the zero set of the barrier---implies that no solution of $\dot{x}=f(x)$ may connect a point with negative value of $V$ to a point with positive value of $V$. This also holds for discrete resets and hence we can handle both the third and fourth part in the same way. The resulting constraints are:

\begin{itemize}
\item $\bigwedge_{(s,s')\in S,s\in\Init} V(p, s)<0$, $\bigwedge_{(s,s')\in S,s'\in\Init} V(p, s')<0$
\item $\bigwedge_{(s,s')\in S,s\in\Unsafe} V(p, s)>0$, $\bigwedge_{(s,s')\in S,s'\in\Unsafe} V(p, s')>0$,
\item $\bigwedge_{(s, s')\in S} V(p, s)>0 \vee V(p, s')<0$
\end{itemize}

We will call the conjunction of these constraints \emph{sampled constraint} and will denote it by $\SampledConstr{S}$. Clearly, this approximation of $C_f$ by $\SampledConstr{S}$ does not lose barrier certificates:
\begin{property}
\[\{ p\mid p\models C_f\}\subseteq \{ p \mid p\models C_S\}\]
\end{property}

Unlike the original constraint $C_f$, the sampled constraint $C_S$ does not contain any quantifier alternation which makes it easier to solve. However, it may have spurious solutions, that is, solutions that do not correspond to a solution of the original constraint and that, hence, do not represent a barrier certificate

In order to handle such a situation, we use the following property: 
\begin{property}
  If $S\subseteq S'$ then $\{ p \mid p\models C_{S'} \} \subseteq \{ p\mid p\models C_{S}\}$.
\end{property}

So adding more segments to $S$ does not weaken the approximation. To actually strengthen the approximation  we use an algorithm based on the principle of counter-example based refinement: The algorithm computes a solution of $C_S$ that we will call \emph{barrier candidate}, checks whether this barrier candidate is spurious, and if yes, generates and adds a counter-example in the form of a new simulation segment that refutes the given barrier candidate. If the barrier candidate is not spurious, we return the vector $p$ which then represents a barrier certificate. 

The resulting algorithm looks as follows:\nopagebreak[3]
\begin{tabbing}\hs\hs\=\hs\hs\=\kill
initialize $S$ with some simulation segments\\
  let $p$ be s.t. $p\models C_S$\\
  \textbf{while} $p\not\models C_f$ \textbf{do}\+\\
     $S\leftarrow S\cup \{(s, s')\}$, where $(s, s')$ is a simulation segment with $p\not\models C_{S\cup \{ (s, s')\}}$\\
     let $p$ be s.t. $p\models C_S$\-\\
\textbf{return} $p$
\end{tabbing}

The algorithm leaves the concrete choice of the barrier candidate and counter-example open. As it is, allowing an arbitrary choice of those objects, it does not work. The main problem is a consequence of the fact that the space of barrier candidates is uncountable. Computing an arbitrary barrier candidate, and then removing this single barrier candidate does, in general, not make enough progress in removing spurious barrier candidates\footnote{Decision procedures for real closed fields can circumvent this problem~\cite{Jovanovic:12}, due to the fact that semi-algebraic sets  possess an algorithmically computable finite cellular decomposition~\cite{Collins:75}. }. Moreover, if the system dynamics $f$ is non-polynomial, it is, in general, not possible to decide the satisfiability test $p\models C_f$ which is the termination condition of the algorithm.

In the next three sections we will design a variant of the above algorithm that overcomes those problems. We will compute a barrier candidate $p$ such that $p\models C_S$ and a counter-example $(s, s')$ with $p\not\models C_{S\cup \{ (s, s')\}}$ that ensure as much progress of the algorithm as possible. That is, the counter-example $(s, s')$ should ensure that $\{ p \mid p\models C_{S\cup\{(s, s')\}}\}$ is significantly smaller than $\{ p\mid p\models C_S\}$ and hence closer to $\{ p \mid p\models C_f\}$. As a side-effect we will also get a termination condition  for the refinement loop that represents a computable and practically reliable replacement for the satisfiability test $p\models C_f$.

\section{Computing a Barrier Candidate}
\label{sec:candidate}

The sampled constraint $\SampledConstr{S}$ can have many solutions. Which one should we choose? Certainly we should prefer non-spurious solutions that is, solutions that also satisfy the original constraint $C_f$. 
Moreover, if a solution turns out to be spurious, removing it should remove as many further spurious solutions as possible. We will work with the assumption, that those objectives will be fulfilled by solutions that are as central as possible in the solution set of the sampled constraint.

For this we replace the inequalities, that can be either satisfied or not, by a finer measure~\cite{Ratschan:02b}. Observing, that the right-hand side of every inequality is zero, we base this measure on the value of the term on the left-hand side: This value measures how strongly a given point $p$ satisfies a greater-than-zero predicate. In the case of a less-than-zero predicate, we can measure this by multiplying the value of the term on the left-hand side by $-1$. Moreover, we replace conjunction by the minimum operator and disjunction by the maximum operator in the style of fuzzy logic. 

The result is the function that assigns to $p$ the value
\[ \min \left\{
\begin{array}{l}
 \min_{s\in I, (s, s')\in S} -V(p, s), \min_{s'\in I, (s, s')\in S} -V(p, s'),\\
 \min_{s\in U, (s, s')\in S} V(p, s),\min_{s'\in U, (s,s')\in S} V(p, s'),\\
 \min_{(s, s')\in S} \max \{V(p, s),-V(p, s')\}  
\end{array}
\right\}.\]
We maximize this function to find points that satisfy the constraint $C_S$ as strongly as possible. 

Now observe that template polynomials $V(p, x)$ of the form $p_0+\sum p_i \vec{x}^i$ are linear in their parameters $p_0, p_1,\dots$. Hence, the result of substituting points $s$ and $s'$ for $x$ in $V(p, x)$ is a linear inequality of the form $a^T p<0$ with $p$ being the parameter vector $(p_0, p_1,\dots)$ and $a$ being a vector of real numbers whose first entry, corresponding to the monomial $p_0$, is the constant~$1$.

For a polynomial template and $\lambda\geq 0$, $V(\lambda p, s)=\lambda V(p, s)$. Hence, also the above function scales in such a way, the corresponding optimization problem is unbounded, and optimization algorithms will usually simply come up with larger and larger values for the vector $p$. In other words, instead of optimizing for our goal of being as much as possible in the solution set of the sampled constraint this formulation optimizes for large parameter values which, in turn, result in large values of $V(p,s)$. 
We avoid this by constraining the (max)-norm of the vector $p$ to not to exceed $1$.

However, even then, minimizing a linear term  $a^T p$ enforces large distance from the boundary of the solution set of $C_S$, if $||a||$ is small, and vice versa. For avoiding this, we normalize the terms, resulting in $\frac{a^T}{||a||_2} p$. This amounts to computation of the Chebyshev center~\cite{Boyd:04}, that is, 
the  center of the largest ball contained in the solution set.\footnote{Note that due to the disjunction, we do not have a polyhedron here. Still, this formulation models the Chebyshev center.}

So we solve the optimization problem
\[ \max_{||p||\leq 1} F_S(p)
\] where $F_S(p)$ is the minimax function above with all linear terms normalized by dividing them with the $2$-norm of their coefficients.

\begin{property}
  \label{prop:5}
  $F_S(p)> 0$  iff   $p\models C_S$ 
\end{property}

Hence, a positive result of the optimization problem gives us a solution of the sampled constraint. By optimizing further, we get solutions that are as central as possible in the solution set of $\SampledConstr{S}$, hence also increasing the chances of finding a solution of the original constraint $C_f$.

\section{Computing a Counter-Example}
\label{sec:cntrxpl}

The solution $p$ of the sampled constraint $C_S$ might be spurious, that is, it might not satisfy the original constraint $C_f$. 
If the computed solution is spurious, we generate a counter-example, that is, a new simulation segment $(s, s')$ s.t. $p$ does not satisfy the strengthened sampled constraint $C_{S\cup \{ (s, s')\}}$. 
However, this constraint should not only refute the computed barrier candidate $p$, but as many further spurious solutions as possible. The techniques from the previous section, that is, maximizing $F_S$ instead of computing an arbitrary solution of $\SampledConstr{S}$, alleviates the problem: It results in a barrier candidate $p$ that is as central as possible in $\{ p \mid p\models C_S\}$ and hence removing this barrier candidate from the set will also significantly shrink this set. However, in addition, we also want to add a simulation segment $(s, s')$ that not only removes the spurious solution $p$ but as many further spurious solutions as possible. 

For this, we again translate the constraint solving problem of finding a counter-example into an optimization problem. However, 
searching for a strong violation of $C_{S\cup \{(s, s')\}}$ by searching for a simulation segment $(s, s')$ s.t. $F_{S\cup \{(s, s')\}}$ is minimal, is an ODE-constrained optimization problem. Such problems are notoriously difficult to solve. In order to avoid this, we work with the original constraint $C_f$, instead. We have a \emph{fixed} barrier candidate $p$, and 
\begin{enumerate}
\item first look for a point $x$ violating the universal quantifier in one of the individual parts of $C_f$, and then
\item compute a counter-example $(s, s')$ by simulating from $x$ using an appropriate simulation length $T$, and disturbance input $d$.
\end{enumerate}

We now analyze the two steps in more detail. Here we call the point $x$ from the first step a \emph{counter-example point}, and call a counter-example also \emph{counter-example segment}.

\subsection{Computing a Counter-Example Point}
\label{sec:ctrxpl-point}

By looking for a counter-example point $x$ violating one of the individual parts of $C_f$ as much as possible we hope to construct a counter-example segment not only for the given spurious candidate $p$, but for as many further spurious candidates as possible. 
Applying the constraint-to-function transformation already described in the previous section to the four parts of the constraint $C_f$, we arrive at the functions
\[
\begin{array}{l}
\min \{ -V(p, x) \mid x\in \Init \},\\
\min \{ V(p, x)\mid x\in\Unsafe\},\\
  \min \{ -(\nabla V(p, x))^T f(x,d) \mid V(p, x)=0, x\in\Inv, d\in D \}, \text{and}\\
  \min \{ \max \{ V(p, x), -V(p,x') \} \mid (x,x')\in\Reset\}
\end{array}
.\] 
However, the third item does not fully correspond to the original intention of the corresponding constraint: Its task is to measure, whether all solutions of the ODE crossing the zero level set $\{ x \mid V(p,x)=0 \}$ do so in the correct direction. This direction should be independent wrt. scaling of $f(x)$ or
 $\nabla V(x)$. In order to normalize those factors, we replace the objective function $-(\nabla V(p, x))^T f(x)$ with the objective function 
\[ -\frac{\nabla V(p,x) }{||\nabla V(p,x) ||}^T \frac{f(x,d)}{||f(x, d||}.\]

As a result, we have optimization problems
\begin{itemize}
\item $\min_{x\in\Init} F_\Init(p,x)$, where $F_\Init(p,x):= -V(p,x)$, 
\item $\min_{x\in\Unsafe} F_\Unsafe(p,x)$, where $F_\Unsafe(x):= V(p,x)$, 
\item $\min_{x\in\Inv, d\in D, V(p,x)=0} F_\nabla(p,x,d)$ where $F_\nabla(p,x,d):=
-\frac{\nabla V(p,x) }{||\nabla V(p,x) ||}^T \frac{f(x,d)}{||f(x,d)||}$, and
\item $\min_{(x,x')\in \Reset} F_r(p, x,x')$ were $F_r(p,x,x'):= \max \{ V(p,x), -V(p, x')\}$.
\end{itemize}

Then we choose the counter-example point as the minimizer corresponding of the minimal result of those four optimization problems. 

Compared to the problem from Section~\ref{sec:candidate}, where the search space is the parameter space, and the state space was discretized, here  $p$ is fixed, and we search in the original state space $\Omega.$

\subsection{Computing a Counter-Example Segment}

If the result of the minimization is negative then we have a point  violating the universal quantifier in one of the individual parts of $C_f$, and we can construct a counter-example segment from this point. For this, we need a disturbance input $d$ and simulation time $T$. For choosing those, we return to the constraint $C_S$ and the corresponding function $F_S$. The following properties ensure the existence of a counter-example segment:

\begin{property}
  \label{prop:3}
  Let $x\in\Init$ with $F_\Init(p,x)\leq 0$. Then for all $x'\in\Omega$, $p\not\models C_{S\cup \{ (x, x')\}}$.  
\end{property}

\begin{proof}
  $F_\Init(p,x)\leq 0$ means $-V(p,x)\leq 0$, that is, $V(p,x)\geq 0$. However, $C_{S\cup \{ (x, x')\}}$ implies $V(p, x)<0$, a contradiction. \qed  
\end{proof}

Note that the proof uses only properties of $V$ at $x$, but does not use the point $x'$ at all. So why does it make sense to use a strictly positive simulation length instead of simply using the solution segment $(x,x)$? The reason is that the counter-example segment $(x, x')$ refutes even more barrier candidates. More specifically, this segment also refutes barrier candidates in the case where $x$ is no counter-example point any more:

\begin{property}
  Let $x\in\Init$ with $F_\Init(p,x)>0$. Then for all  $x'\in\Omega$ with $V(p,x')>0$, $p\not\models C_{S\cup \{ (x, x')\}}$.  
\end{property}


\begin{proof}
  If $F_\Init(p,x)>0$ then $V(p,x)<0$. Moreover, the third part of $C_S$ is $V(p, x)>0 \vee V(p, x')<0$ which contradicts $V(p,x)<0$, $V(p,x')>0$ and hence refutes $p$ with $V(p, x')<0$. \qed
\end{proof}

The premises $F_\Init(p,x)>0$ and $V(p,x')>0$ can only be simultaneously fulfilled for simulations of non-zero length, which motivates such a choice. Due to similar reasons it does not make sense to use backward simulations from points $x\in\Init$. However, for points $x\in \Unsafe$, the same arguments hold in reverse time. Especially, we have the following dual of Property~\ref{prop:3}:


\begin{property}
  \label{prop:4}
Let $x\in\Unsafe$ with $F_\Unsafe(p,x)<0$. Then for all $x'$ with $x'\in\Omega$, $p\not\models C_{S\cup \{ (x', x)\}}$.  
\end{property}

For counter-example points corresponding to $F_\nabla$ and $F_r$ we have similar properties. However, in the case of $F_\nabla$, even for refuting the given barrier candidate $p$, a single point does not suffice and we need a counter-example segment of non-zero length:

\begin{property}
\label{prop:1}
Let $x\in \Omega$, $d$ be a disturbance input such that $V(p,x)=0$ and $F_\nabla(p,x,d)<0$. Then there are $T^*_->0, T^*_+>0 $ s.t. for all $0<T_-\leq T^*_-, 0<T_+\leq T^*_+$, $x^-$, $x^+$ with $x^-\xrightarrow{d,T^*_-} x\xrightarrow{d, T^*_+} x^+$, $p\not\models C_{S\cup \{ (x^-, x^+)\}}$.  
\end{property}

\begin{proof}
  $V(p, x)=0$ and $F_\nabla(p,x, d)<0$ means that $(\nabla V(x))^T f(x,
  d)<0$. Let $y(t)$ be a solution of the ODE defining the continuous dynamics in the mode of $x$ s.t. $y(0)=x$.  Then the derivative of $V(y(t))$, with $y(t)$ being the solution   as a function in $t$ is strictly negative at $y(t)$. This derivative is continuous and hence  there is there is $T^*_+>0, T^*_->0$ s.t. for all $0<T_-\leq T^*_-, 0<T_+\leq T^*_+$, $x^-$, $x^+$ with $x^-\xrightarrow{d,T^*_-} x\xrightarrow{d, T^*_+} x^+$,
$V(x^-)<0$ and $V(x^+)>0$. However, $C_{S\cup \{ (x^-,x^+)\}}$ implies $V(x^-)>0$ or $V(x^+)<0$, a contradiction. \qed
  
\end{proof}

\begin{property}
\label{prop:2}
Let $x, x'\in\Omega$ with $F_r(p,x,x')<0$. Then for all disturbance inputs~$d$ there are $T^*_-, T^*_+>0$ s.t. for all $0\leq T_-\leq T^*_-$, $0\leq T_+\leq T^*_+$, $x^-$, $x^+$ with $x^-\xrightarrow{d,T^*_-} x$, $x'\xrightarrow{d, T^*_+} x^+$, $p\not\models C_{S\cup \{ (x^-, x^+)\}}$.  
\end{property}

\begin{proof}
Since  $F_r(p,x,x')<0$,  $\max\{ V(p,x), -V(p, x')\}<0$, and hence  $V(p,x)<0$ and $V(p,x')>0$.
So there is $T^*_+>0, T^*_->0$ s.t. for all $0\leq T_-\leq T^*_-, 0 \leq T_+\leq T^*_+$,
$x^-$, $x^+$ with $x^-\xrightarrow{d,T^*_-} x$, $x'\xrightarrow{d, T^*_+} x^+$,
$x^-<0$ and $x^+>0$.  However, $C_{S\cup \{ (x^-, x^+) \}}$ implies $V(x^-)>0$ or $V(x^+)<0$, a contradiction. \qed
\end{proof}

Summarizing, for a counter-example point resulting from $F_I$, we do a forward simulation, for a counter-example point resulting from $F_U$ a backward simulation, for a counter-example point resulting from $F_\nabla$ we simulate in both directions, and for a pair of counter-example points $(x,x')$ resulting from $F_r$ we simulate backward from $x$ and forward from $x'$.


However, in all properties above, the disturbance inputs and simulation lengths are universally quantified, and the question is, which one to choose. Even more, in the case of Property~\ref{prop:3} the proof is fully independent of the simulation length. So we need more considerations to choose those values. 
The choice should ensure that the added solution segment refutes not only the computed candidate $p$ that turned out to be spurious, but as many further spurious candidates as possible. For this, we add a segment $(s, s')$ that refutes the spurious barrier candidate $p$ as \emph{much} as possible---expecting that this also refutes a large neighborhood of $p$. As discussed in Section~\ref{sec:candidate}, we measure 
fulfillment of $C_S$ using the function $F_S$ which is~\footnote{up to---here irrelevant---scaling}
\[ \min \left\{
\begin{array}{l}
 \min_{s\in I, (s, s')\in S} -V(p, s), \min_{s'\in I, (s, s')\in S} -V(p, s'),\\
 \min_{s\in U, (s, s')\in S} V(p, s),\min_{s'\in U, (s,s')\in S} V(p, s'),\\
 \min_{(s, s')\in S} \max \{V(p, s),-V(p, s')\}  
\end{array}
\right\}\]
and hence we  add a segment $(s, s')$ to $S$ that minimizes this function. \Comment{\footnote{add quantitative versions of the above properties here? later, certainly not now}}

In the case of the first line, $s$ is fixed, and we compute $s'$. In the case of the second line, $s'$ is fixed and we compute $s$. The third line results from either $F_\nabla$ or $F_r$, and in both cases we compute both end-points.
Here, in all cases where we compute $s$, the sign of $V(p, s)$ is positive, and in all cases where we compute $s'$, the sign of $V(p, s')$ negative. Hence we want to  minimize $V(p, s)$ and maximize $V(p,s')$. Note that here we compute $s$ using a backward simulation and $s'$ using a forward simulation.

For ensuring minimization of $V(p, s)$ and maximization of $V(p,s')$, we choose the disturbance input $d$ that maximizes $\nabla V(p,s) f(s,d)$ ($\nabla V(p,s') f(s',d)$, respectively). Moreover, we simulate as long as $V(p,s)$ decreases ($V(p,s')$ increases, respectively) that is, until $\nabla V(p,s) f(s,d)$ ($\nabla V(p,s') f(s',d)$, respectively) is zero. We also terminate simulations that hit an initial point in the backward direction, an unsafe point in the forward direction, or a point $x$ with $x\not\in\Inv$.



For a counter-example point $x$ to the transversality condition the value $\nabla V(p,x) f(x,d)$  is always positive. In the other cases, $F_\Init$, $F_\Unsafe$, $F_r$, it might be zero or negative which means that a simulation from this point will not result in a desired increase. In such cases, we simply use a simulation time~$0$ which results in the according endpoint of the simulation segment to be equal to the counter-example point.

In the rest of the paper, we will write $\omega(x)$ for the endpoint of a simulation starting from $x$ with $d$ and $T$ chosen as described above, and $\alpha(x)$ the starting point of a backward simulation from $x$ with $d$ and $T$ chosen as described above.

Here we have a chicken-and-egg problem: For simulation we need a barrier candidate and vice versa. This can be resolved by either starting with trivial simulations or with a trivial barrier candidate. The current solution is to start with simulations of \emph{fixed} length from every box vertex.

\section{Resulting Algorithm}
\label{sec:algorithm}

\begin{tabbing}\hs\hs\=\hs\hs\=\kill
initialize $S$ with some simulation segments\\
  $(cand, cntrxpl)\leftarrow check(S)$\\
  \textbf{while} $\neg [cand=\bot \vee ctrxpl = \emptyset]$ \textbf{do}\+\\
     $S\leftarrow S\cup cntrxpl$\`\\
     $(cand, cntrxpl)\leftarrow check(S)$\-\\
\textbf{if} $cand=\bot$ \textbf{then}
  \textbf{return} ``no barrier found''\\
  rigorously verify $cand$\`\emph{optional verification step}\\
\textbf{return} $cand$
\end{tabbing}

\begin{tabbing}\hs\hs\=\hs\hs\=\kill
\textbf{subalgorithm} check(S): returns barrier candidate and counter-example\\
  let $p$ be s.t. $F_S(p)$ is as large as possible\`\emph{compute a barrier candidate}\\
  \textbf{if} $F_S(p)\leq 0$ \textbf{then}
    \textbf{return} $(\bot, \emptyset)$\`\emph{no barrier candidate found}\\
    let $x_{\Init}\in \Init$ be s.t. $F_\Init(p,x_I)$ is as small as possible\\
    let $x_{\Unsafe}\in \Unsafe$ be s.t. $F_\Unsafe(p,x_U)$ is as small as possible\\
    let $x_{\Barrier}\in\Inv, d\in D$ be s.t. $V(p, x_\Barrier)=0$ and  $F_\Barrier(p,x_\Barrier,d)$ is as small as possible\\
    let $x_{r}, x_{r}'\in \Reset$ be s.t. $F_r(p,x_r, x_r')$ is as small as possible\\
    $v\leftarrow \min \{ F_I(p, x_\Init), F_U(p, x_\Unsafe), F_\Barrier(p, x_\Barrier, d), F_r(p, x_r)\}$\\
    \textbf{if} $v \geq 0$ \textbf{then}
    \textbf{return} $(p, \emptyset)$\`\emph{no counterexample found}\\
    \textbf{if} $v=F_\Init(p, x_\Init)$ \textbf{then} \textbf{return} $(p, \{ (x_\Init, \omega(x_\Init))\})$\\
    \textbf{else} \textbf{if} $v=F_\Unsafe(p, x_\Unsafe)$ \textbf{then} \textbf{return} $(p, \{ (\alpha(x_\Unsafe), x_\Unsafe)\})$\\
    \textbf{else} \textbf{if} $v=F_\Barrier(p, x_\Barrier,d)$ \textbf{then} \textbf{return} $(p, \{ (\alpha(x_\Barrier), \omega(x_\Barrier))\})$\\
    \textbf{else} \textbf{if} $v=F_r(p, x_r)$ \textbf{then} \textbf{return} $(p, \{ (\alpha(x_r), \omega(x_r'))\})$


\end{tabbing}


Note that here we only need values for which the objective functions are large (small, respectively). We do \emph{not} insist on a lower bound of the minimization problem (upper bound on the maximization problem, respectively), let alone a decision procedure. This allows the use of various heuristic optimization techniques~\cite{Locatelli:13} that even can be applied in cases where finding a precise optimum is impossible due to non-decidability issues, for example, due to non-polynomial system dynamics $f$ occurring in  $F_\nabla$.

Also observe that the optimization of $F_S(p)$ is a search problem  of the parameter space dimension~$k$, and the computation of $x_I$, $x_U$, $x_\nabla$, and $x_\Reset$ is a search problem of the state space dimension~$n$. In contrast to that, directly solving original constraint $C_f$ is a problem in dimension $n+k$. \Comment{\footnote{disturbance inputs?}}

The final step of rigorously verifying the barrier candidate, that is, verifying $p\models C_f$, is a problem in state space dimension $n$, as well. Due to the strategy of optimizing for a barrier candidate, the computed candidate will usually satisfy $C_f$ robustly. Hence, even in undecidable cases, this allows the application of procedures that exploit robustness~\cite{Ratschan:02f}.

\Comment{\footnote{include figures illustrating algorithm behavior?}}

\section{Implementation}
\label{sec:implementation}

In the section, we show how the optimization problems and the final verification step of the algorithm from the previous section can be solved in practice.


As described in Section~\ref{sec:candidate}, $F_S(p)$ is linear in $p$. However, it contains a $\min$/$\max$ alternation which is beyond the capabilities of usual numerical optimization algorithms. The key to solving this constraint is the observation that the $\min$/$\max$ operators occurring within $F_S(p)$ are \emph{finite}. Hence the optimization problem can be rewritten to the following constrained optimization problem: 
Maximize $\delta$ under\Comment{\footnote{This seems to correspond to linear programming support vector machines~\cite{Zhou:02}. There seem to be further possibilities! \cite{Abe:10}}}
\[
\begin{array}{l}
\bigwedge_{(s,s')\in S, \Init(s)} -V(p, s)\geq \delta,
\bigwedge_{(s,s')\in S, \Init(s')} -V(p, s')\geq \delta,\\
\bigwedge_{(s,s')\in S, \Unsafe(s)} V(p, s)\geq \delta,
\bigwedge_{(s,s')\in S, \Unsafe(s')} V(p, s')\geq \delta, \text{ and}\\
\bigwedge_{(s, s')\in S} V(p, s)\geq \delta \vee -V(p, s')\geq \delta.
\end{array}\]
This is  an optimization modulo theory~\cite{Nieuwenhuis:06,Sebastiani:15a} problem in the theory LRA (linear real arithmetic). 

For minimizing $F_{\Init}(p, x)$, $F_{\Unsafe}(p, x)$, $F_{\Barrier}(p, x)$, and $F_{r}(p, x)$, one can use classical numerical optimization~\cite{Nocedal:99}. Since such methods do local search, they may run into local, but non-global optima. To search for global solutions one can start several optimization runs from random starting points which is also known under the term multi-start~\cite{Marti:03}. Note that this is trivial to parallelize efficiently. The $\min$/$\max$ alternation in $F_r$ can be handled as above.

For the final rigorous verification step, one can use a simple branch-and-bound approach, evaluating the terms $V(p,x)$ using interval arithmetic~\cite{Moore:09}, checking the inequalities of Definition~\ref{def:1} on the resulting intervals, and using splitting to tighten the bounds, if necessary. 

\section{Computational Experiments}
\label{sec:experiments}

\newcommand{\Unused}[1]{}

We did experiments with a prototype implementation of the method described so far. The prototype requires the state space, set of initial states and the set of unsafe states to have the shape of a hyper-rectangle. We initialize the set $S$ by forward simulations from all vertices of the initial hyper-rectangle and backward simulations from all vertices of the unsafe hyper-rectangle. Due to this initialization, our prototype implementation does not check barrier candidates for violations of the first two conditions of Definition~\ref{def:1}, and indeed, even without such a check, the computed barriers do not violate those conditions.

For each example, we set  the lengths of all simulations manually to a certain constant $\sigma$ that we show below. Moreover, we cancel simulations that leave a bloated version of the state space. Here, we simply bloat each interval bound of $\Omega$ by a certain percentage from its distance from the interval center:
$bloat([\underline{a}, \overline{a}])= [\frac{\underline{a}+\overline{a}}{2}-b(\frac{\underline{a}+\overline{a}}{2}-\underline{a}), \frac{\underline{a}+\overline{a}}{2}+b(\overline{a}-\frac{\underline{a}+\overline{a}}{2})]= [ \frac{(1+b)\underline{a}+(1-b)\overline{a}}{2}, \frac{(1-b)\underline{a}+(1+b)\overline{a}}{2}]$. 
In our experiments, we use $b=1.1$.




The examples that we used are all purely continuous, without any hybrid behavior: 

\begin{enumerate}
\Unused{
\item 

\[
\begin{array}{l}
\dot{x}=-x-y\\
\dot{y}=x-y
\end{array}
\]

$\Omega= [-10,10]\times[-10,10]$, $\Init= [-1,1]\times[-1,1]$, $\Unsafe= [5,10]\times [5,10], \sigma=0.5$
\item 

\[
\begin{array}{l}
\dot{x}=-0.5x\\
\dot{y}=0.5y
\end{array}
\]
$\Omega= [-10,10]\times[-10,10]$, $\Init= [-1,1]\times[-1,1]$, $\Unsafe= [-2.6,-0.6]\times [-6.5,-4.5], \sigma=1.0$
\item 

\[
\begin{array}{l}
\dot{x}=y\\
\dot{y}=-x+x^3-y
\end{array}
\]
$\Omega= [-10,10]\times[-10,10]$, $\Init= [-1,1]\times[-1,1]$, $\Unsafe= [-2.6,-0.6]\times [-6.5,-4.5], \sigma=0.5$

Prajna~\cite{Prajna:04}, Example~1, with $p=1$, disks changed to boxes, originally Khalil (3rd ed. p. 315), Prajna needs a quartic barrier! quadratic barrier does not suffice

\item 
a standard ODE modeling a pendulum with normalized
parameters (e.g., Kapinski et al.~\cite{Kapinski:14}, Example 1)
\[
\begin{array}{l}
\dot{x}=y\\    
\dot{y}= - \sin x - y 
\end{array}
\]
Here the variable $x$ models the angle of the pendulum, and $y$ models angular speed.

$\Omega= [-10,10]\times[-10,10]$, $\Init= [-1,1]\times[8,10]$, $\Unsafe= [-10,10]\times [-10,-5], \sigma=0.5$
}
\item 
a standard ODE modeling a pendulum with normalized
parameters (e.g., Kapinski et al.~\cite{Kapinski:14}, Example 1), where the variable $x$ models the angle of the pendulum, and $y$ models angular speed.
\[
\begin{array}{l}
\dot{x}=y\\    
\dot{y}= - \sin x - y 
\end{array}
\]

$\Omega= [-10,10]\times[-10,10]$, $\Init= [-10,10]\times[8,10]$, $\Unsafe= [-10,10]\times [-10,-5], \sigma=0.5$

\Unused{
\item 
\cite{Bouissou:14}: Example~1, 
\[
\begin{array}{l}
  \dot{x} = x + y\\
  \dot{y} = xy-\frac{1}{2} y^2
\end{array}
\]
$\Omega= [-10,0]\times[0,10]$, $\Init= [0,0.2]\times[-0.2,0]$, $\Unsafe= [-2,-1]\times [-2,-1], \sigma=0.2$
\item 
same dynamics as previous example , but (adapted from arXiv version of paper)

$\Omega= [-10,0]\times[0,10]$, $\Init= [-1.5,-1.0]\times[1, 1.5]$, $\Unsafe= [-2.7,-2.3]\times [0.5, 1.0], \sigma=0.2$
\item 
example that apparently unsafe but chapoutot says barrier found

\[
\begin{array}{l}
\dot{x}= -x + xy\\ 
\dot{y}= -y\\
\end{array}
\]

\item 
 \cite[Example 4.1]{Papachristodoulou:05}, \cite[Example 7]{Bouissou:14}

\[
\begin{array}{l}
  \dot{x}=y\\
  \dot{y}= - \frac{x+y}{\sqrt{1+(x+y)^2}} 
\end{array}
\]

$\Omega= [-10,10]\times[-10,10]$, $\Init= [-0.2,0.2]\times[-0.2,0.2]$, $\Unsafe= [3.3,3.7]\times [0.3, 0.7], \sigma=1$

\item 
 same as previous, variant~\cite[Example 7]{Djaballah:15}

$\Omega= [-1000,1000]\times[-100,100]$, $\Init= [-0.5,0.5]\times[-0.5,0.5]$, $\Unsafe= [2.5,3]\times [-0.5, 0]$, $\sigma=1$
\item 
\cite[Example 5]{Djaballah:15} 

\[
\begin{array}{l}
  \dot{x} = y + (1 - x^2 - y^2)x + \ln (x^2 + 1)\\
  \dot{y} = - x + (1 - x^2 - y^2)y + \ln (y^2 + 1)
\end{array}
\]
$\Omega= [-10,10]\times[-10,10]$, $\Init= [1,3]\times[-1.5,3.0]$, $\Unsafe= [-0.62,-0.58]\times[0.98,1.02], \sigma=1$
}

\item 
dynamics from~\cite[Example 5]{Djaballah:15} 

\[
\begin{array}{l}
  \dot{x} = y + (1 - x^2 - y^2)x + \ln (x^2 + 1)\\
  \dot{y} = - x + (1 - x^2 - y^2)y + \ln (y^2 + 1)
\end{array}
\]

$\Omega= [-5,5]\times[-5,5]$, $\Init= [1,3]\times[-1.5,3.0]$, $\Unsafe= [-3,-0.6]\times[1,3], \sigma=1$

\Unused{
\item 
\cite[Example 5]{Bouissou:14}
\[
\begin{array}{l}
  \dot{x}= y\\
  \dot{y}= \frac{\sqrt{1 + y^2}}{25 - x}
\end{array}
\]
$\Omega= [-100,100]\times[-1,-1]$, this is a mistake! recompute!

\item  
trivial example
\item  
\cite[Example 1]{Zeng:16} 
\[
\begin{array}{l}
  \dot{x}= - x + 2 x^2 y\\
  \dot{y}= - y
\end{array}
\]
$\Omega= [-2,2]\times[-2,2]$, $\Init= [-0.5, 0.5], [0.5, 1.5]$, $\Unsafe= [1.2, 1.8]\times [1.2, 1.8], \sigma=1$
\item  
\cite{Lin:16}; originally from \cite{Chesi:09} apparently has no quadratic barrier
\[
\begin{array}{l}
\dot{x}= y\\
\dot{y}= - y - 10 \sin x
\end{array}
\]

$\Omega= [-2,2]\times[-2,2]$, $\Init= [1.2, 1.8], [-0.3, 0.3]$, $\Unsafe= [-1.3, -0.7]\times [0.7, 1.3], \sigma=0.2$
}
\item 
a standard Lorenz system \cite{Vanecek:96}, see also~\cite[Example 7]{Djaballah:17}

\[
\begin{array}{l}
  \dot{x} = 10  (y - x)\\
  \dot{y} = x  (28 - z)- y\\
  \dot{z} = x  y - \frac{8}{3} z
\end{array}
\]

$\Omega= [-20, 20]\times[-20,0]\times [-20,20]$, 
$\Init= [-14.8, -14.2]\times [-14.8, -14.2]\times [12.2, 12.8]$, 
$\Unsafe= [-16.8, -16.2]\times [-14.8, -14.2]\times [2.2, 2.8]$, $\sigma=0.1$

\item 
composition of trivial dynamics (variable $x_1$) and pendulum (variables $x_2$ and $x_3$)
\[
\begin{array}{l}
\dot{x}_1= 1\\
\dot{x}_2= x_3\\
\dot{x}_3= - 10 \sin x_2 - x_3
\end{array}\]

$\Omega= [-10,10]^3$, $\Init= [9, 10]\times[-10,10]^2$, $\Unsafe= [-10,-9]\times [-10,10]^2, \sigma=0.1$

\item\label{xpl:scale} 
 scalable example, manually constructed
\[
\begin{array}{l}
\dot{x}_1= 1+\frac{1}{l}(\sum_{i\in\{1,\dots,l\}} x_{i+1}+x_{i+2}))\\
\dot{x}_{2}= x_{3}\\
\dot{x}_{3}= - 10 \sin x_{2} - x_{2}\\
\dots\\
\dot{x}_{2l}= x_{2l+1}\\
\dot{x}_{2l+1}= - 10 \sin x_{2l} - x_{2}
\end{array}\]

$\Omega= [-10,10]^{2l+1}$, $\Init= [9, 10]\times[-10,10]^{2l}$, $\Unsafe= [-10,-9]\times [-10,10]^{2l}, \sigma=0.1$, with $l=100$
\item 
same as Example~\ref{xpl:scale}, but $l=2$

\item 
same as Example~\ref{xpl:scale}, but $l=3$
\item 
same as Example~\ref{xpl:scale}, but $l=4$

\end{enumerate}

All experiments were executed on a notebook with Intel(R) Core(TM) i7-5600U CPU @ 2.60GHz and running Ubuntu Linux 16.10. 
For simulation we used the software package CVODE version 2.5.0 from the SUNDIALS suite of solvers. For optimizing $F_S(p)$ we use the tool OptiMathSAT~\cite{Sebastiani:15}. For minimizing $F_{\Init}(p, x)$, $F_{\Unsafe}(p, x)$, and $F_{\Barrier}(p, x)$
 we use the function sqp  from the software package GNU Octave 4.0.3 which implements the optimization method of sequential quadratic programming. We globalized this method by multi-start with 16 local optimization runs. For the final rigorous verification step, we use our software \textsf{RSolver} (\url{http://rsolver.sourceforge.net}) which extends a basic interval branch-and-bound method with interval constraint propagation.



\begin{table}[htb]
  \centering
{\large
\begin{tabular}{|r|r|c|r|D{.}{.}{2}|D{.}{.}{2}|D{.}{.}{2}|D{.}{.}{2}|}\hline
&dim &templ&iter&\multicolumn{1}{c|}{simulation}&\multicolumn{1}{c|}{candidate}&\multicolumn{1}{c|}{counter-example}&\multicolumn{1}{c|}{verif}\\\hline
1 & 2& Q&10&0.24&1.2&8.21&0\\
2&2 & Q&5&0.11&0.25&5.7&0.41\\
3&3 &T&10&0.3&1.01&17.03&0\\
4&3& L&1&0.02&0&1.07&0\\
5&3&L&1&0.01&0.01&1.21&0\\
6&5&L&1&0.14&0.36&3.51&0\\
7&7&L&1&1.06&7.38&7.67&0\\
8&9&L&1&15.81&1340.6&19.76&0.01\\
\hline
\end{tabular}  }
  \caption{Results of Experiments}
  \label{tbl:results}
\end{table}

We list the results in Table~\ref{tbl:results}. Here, the column ``dim'' denotes the problem dimension and ``templ'' denotes one of the following templates:
\begin{description}
\item[Q: ] $p_0+ p_1x^2+p_2 xy + p_3 y^2 + p_4 x + p_5 y$
\item[T: ] $p_0 + p_1 x^2 + p_2 x + p_3 z$
\item[L: ] the linear template $p_0 + p_1 x_1 +\dots+p_n x_n$ with $n$ being the state space dimension
\end{description}
Moreover, the column ``iter'' denotes the number of iterations of the refinement loop. Further columns denote the 
the time spent in simulation, computation of a barrier candidate, computation of a counter-example, and verification. The time unit are seconds. 

As can be seen, in all cases, the computed barrier could be rigorously verified. Moreoever, the time needed to do so is negligible. The whole method scales to higher-dimensional examples, but as the problem dimension increases, the optimization module theory solver used to compute a barrier candidate is increasingly becoming a bottleneck. Note that we used the solver as a black box, with the original parameter settings.

To ensure verifiability of our results, we list the computed barriers: {\small
\begin{enumerate}
\item $0.118462553528y^2-0.011722981249xy-0.709542580128y-0.0550927673883x^2-0.0586149062452x-1$
\item $0.408692986165y^2-0.386033509251xy-0.227005969996y+0.0866893912879x^2-0.925807829028x-1$
\item $(-z)+0.0862165171738x^2+0.406513973333x-0.668459116412$

\item $0.12774317671-x_1$
\item $6.94919072662\times 10^{-4}x_3+7.29701934574\times 10^{-4}x_2-x_1+0.127740909365$
\item $0.00298446742425x_5-0.00705872836204x_4-0.00693382587388x_3+0.00295825595803x_2-x_1+0.100721787174$
\item $0.00567387721155x_7+0.00131139026963x_6+0.00409187476431x_5-0.00293955884622x_4-0.00148234438362x_3+0.0102405191466x_2-x_1+0.0693868524466$
\item $0.00474371409319x_9+6.04082564889\times 10^{-4}x_8+0.00539357982978x_7-3.62914727064\times 10^{-5}x_6-0.00305191611365x_5+0.00234411670971x_4+0.00308900495946x_3+0.00766513576991x_2-x_1+0.0526159036023$
\end{enumerate}}

\section{Related Work}
\label{sec:related-work}


The original method for computing barrier certificates~\cite{Prajna:04} was based on sums-of-squares programming~\cite{Parrilo:03}. Since then, various further methods for computing barrier certificates and inductive invariants of polynomials systems have been designed~\cite{Sankaranarayanan:08a,Gulwani:08,Kong:13,Yang:15,Yang:16,Ghorbal:16}.  \Comment{those papers (all of them?) relax the barrier conditions, we do not; discuss this, maybe even in introduction}

To the best of our knowledge, there is only one method capable of computing barrier certificates for non-polynomial systems~\cite{Djaballah:17}. The method is not based on simulation but uses interval-based constraint solving techniques, in a similar way as we do in the final verification step, and in a similar way as the algorithm implemented in \textsc{RSolver}~\cite{Ratschan:02f}. This restricts the method to systems where such techniques are available, which corresponds to those systems, where our algorithm  can do the final verification step. The method applies branching to \emph{both} the state and parameter space, whereas our algorithm, at a given time,  always searches only in one of the two. Instead of our method for computing barrier candidates, the method 
guesses barrier certificates by simply trying midpoints of intervals which can be very efficient if this guess happens to be lucky, but very inefficient, if not. Especially, if the midpoint of the user-provided parameter space already happens to be a barrier certificate, then the method succeeds without any search. Unfortunately, the paper does not give any information on the computed barriers, which makes comparison difficult. 

The approach to generalize or learn system behavior from simulations has been used before for computing Lyapunov functions~\cite{Kapinski:14,Kapinski:15}\Comment{\footnote{add a little bit more on the difference, they also mention barriers}} and for computing the region of attraction \cite{Kozarev:16}. 
Simulations can also be used to directly verify system behavior~\cite{Girard:06,Fainekos:06,Donze:07,Fan:16}. For an overview of simulation-based approaches to systems verification see Kapinski et al.~\cite{Kapinski:16}.




In software verification, the usage of test runs was shown to be useful in the computation of inductive invariants~\cite{Gupta:13,Sharma:13,Garg:14}. However, the problem and solution are quite different from what we have here due to the discrete nature of both time and data types occurring in computer programs.





Our algorithm can also be interpreted as an online machine learning~\cite{Mohri:12} process that learns a barrier certificate from simulations, querying for new simulations to improve the barrier certificate. Moreover, the samples reachable from an initial state or leading to an unsafe state  can be interpreted as positive and negative examples. However, here we do \emph{not} have a classification problem due to the third property of Definition~\ref{def:1}.


The algorithm in this paper adapts  counter-example guided inductive synthesis (CEGIS) that in its original form~\cite{SolarLezama:06} solves discrete constraints with a quantifier prefix $\exists\forall$ to solving certain continuous constraints with such a prefix. However, CEGIS does not work with simulations but only with counter-example points, and it uses constraint solving instead of optimization to find candidates and counter-examples. Some simulation based approaches for computing Lyapunov functions~\cite{Kapinski:14,Kapinski:15} also can be interpreted as continuous versions of CEGIS.

\Comment{applications~\cite{Yang:17}: uses barrier certificates to compute reach set approximation which is then used to switch between controller optimized for performance and safe controller, Jha/Seshia: Theory of Formal Synthesis via Inductive Learning }

\section{Conclusion}
\label{sec:conclusion}

In this paper, we have presented an approach for synthesizing barrier certificates from system simulations. The resulting method is able to compute barrier certificates for ODEs that have been out of reach for such methods so far. 

In the future we will increase the usability of the method by automatizing the choice of the used template. We will also combine the method with falsification methods~\cite{Kuratko:14} that search for ODE solutions that lead from an initial to an unsafe state. In such a combined method, falsification should exploit the result of failed attempts at computing a barrier certificate and vice versa.

\bibliographystyle{abbrv}
\bibliography{sratscha}

\newcommand{\SortNoop}[1]{}
\begin{thebibliography}{10}

\bibitem{Abe:10}
S.~Abe.
\newblock {\em Variants of Support Vector Machines}, pages 163--226.
\newblock Springer London, London, 2010.

\bibitem{Bournez:08}
O.~Bournez and M.~L. Campagnolo.
\newblock A survey on continuous time computations.
\newblock In S.~Cooper, B.~L{\"o}we, and A.~Sorbi, editors, {\em New
  Computational Paradigms}, pages 383--423. Springer New York, 2008.

\bibitem{Boyd:04}
S.~Boyd and L.~Vandenberghe.
\newblock {\em Convex Optimization}.
\newblock Cambridge University Press, 2004.

\bibitem{Caviness:98}
B.~F. Caviness and J.~R. Johnson, editors.
\newblock {\em Quantifier Elimination and Cylindrical Algebraic Decomposition}.
\newblock Springer, Wien, 1998.

\bibitem{Collins:75}
G.~E. Collins.
\newblock Quantifier elimination for the elementary theory of real closed
  fields by cylindrical algebraic decomposition.
\newblock In {\em Second GI Conf. Automata Theory and Formal Languages},
  volume~33 of {\em LNCS}, pages 134--183. Springer Verlag, 1975.
\newblock Also in~\cite{Caviness:98}.

\bibitem{Djaballah:15}
A.~Djaballah, A.~Chapoutot, M.~Kieffer, and O.~Bouissou.
\newblock Construction of parametric barrier functions for dynamical systems
  using interval analysis.
\newblock arXiv:1506.05885v1, 2015.

\bibitem{Djaballah:17}
A.~Djaballah, A.~Chapoutot, M.~Kieffer, and O.~Bouissou.
\newblock Construction of parametric barrier functions for dynamical systems
  using interval analysis.
\newblock {\em Automatica}, 78:287--296, 2017.

\bibitem{Donze:07}
A.~Donz{\'e} and O.~Maler.
\newblock Systematic simulation using sensitivity analysis.
\newblock In A.~Bemporad, A.~Bicchi, and G.~Buttazzo, editors, {\em HSCC 07},
  volume 4416 of {\em LNCS}, pages 174--189. Springer, 2007.

\bibitem{Fainekos:06}
G.~E. Fainekos, A.~Girard, and G.~J. Pappas.
\newblock Temporal logic verification using simulation.
\newblock In {\em International Conference on Formal Modeling and Analysis of
  Timed Systems}, pages 171--186. Springer, 2006.

\bibitem{Fan:16}
C.~Fan, B.~Qi, S.~Mitra, M.~Viswanathan, and P.~S. Duggirala.
\newblock Automatic reachability analysis for nonlinear hybrid models with
  {C2E2}.
\newblock In S.~Chaudhuri and A.~Farzan, editors, {\em Computer Aided
  Verification: 28th International Conference, CAV 2016, Toronto, ON, Canada,
  July 17-23, 2016, Proceedings, Part I}, pages 531--538, Cham, 2016. Springer
  International Publishing.

\bibitem{Garg:14}
P.~Garg, C.~L{\"o}ding, P.~Madhusudan, and D.~Neider.
\newblock {ICE}: A robust framework for learning invariants.
\newblock In {\em International Conference on Computer Aided Verification},
  pages 69--87. Springer, 2014.

\bibitem{Ghorbal:16}
K.~Ghorbal, A.~Sogokon, and A.~Platzer.
\newblock A hierarchy of proof rules for checking positive invariance of
  algebraic and semi-algebraic sets.
\newblock {\em Computer Languages, Systems \& Structures}, 47:19--43, 2017.

\bibitem{Girard:06}
A.~Girard and G.~Pappas.
\newblock Verification using simulation.
\newblock In J.~Hespanha and A.~Tiwari, editors, {\em HSCC'06}, volume 3927 of
  {\em LNCS}, pages 272--286. Springer, 2006.

\bibitem{Gulwani:08}
S.~Gulwani and A.~Tiwari.
\newblock Constraint-based approach for verification and synthesis of hybrid
  systems.
\newblock In {\em Computer Aided Verification}, number 5123 in LNCS, pages
  190--203. Springer, 2008.

\bibitem{Gupta:13}
A.~Gupta, R.~Majumdar, and A.~Rybalchenko.
\newblock From tests to proofs.
\newblock {\em International Journal on Software Tools for Technology
  Transfer}, 15(4):291--303, 2013.

\bibitem{Hainry:08}
E.~Hainry.
\newblock Reachability in linear dynamical systems.
\newblock In A.~Beckmann, C.~Dimitracopoulos, and B.~Löwe, editors, {\em Logic
  and Theory of Algorithms}, volume 5028 of {\em Lecture Notes in Computer
  Science}, pages 241--250. Springer Berlin Heidelberg, 2008.

\bibitem{Henzinger:98}
T.~A. Henzinger, P.~W. Kopke, A.~Puri, and P.~Varaiya.
\newblock What's decidable about hybrid automata.
\newblock {\em Journal of Computer and System Sciences}, 57:94--124, 1998.

\bibitem{Jovanovic:12}
D.~Jovanovi\'{c} and L.~de~Moura.
\newblock Solving non-linear arithmetic.
\newblock In {\em Automated Reasoning - 6th International Joint Conference,
  IJCAR 2012, Manchester, UK, June 26-29, 2012. Proceedings}, volume 7364 of
  {\em Lecture Notes in Computer Science}, pages 339--354. Springer, 2012.

\bibitem{Kapinski:16}
J.~Kapinski, J.~V. Deshmukh, X.~Jin, H.~Ito, and K.~Butts.
\newblock Simulation-based approaches for verification of embedded control
  systems: An overview of traditional and advanced modeling, testing, and
  verification techniques.
\newblock {\em IEEE Control Systems}, 36(6):45--64, 2016.

\bibitem{Kapinski:15}
J.~Kapinski, J.~V. Deshmukh, X.~Jin, H.~Ito, and K.~R. Butts.
\newblock Simulation-guided approaches for verification of automotive
  powertrain control systems.
\newblock In {\em American Control Conference, {ACC} 2015, Chicago, IL, USA,
  July 1-3, 2015}, pages 4086--4095, 2015.

\bibitem{Kapinski:14}
J.~Kapinski, J.~V. Deshmukh, S.~Sankaranarayanan, and N.~Arechiga.
\newblock Simulation-guided {L}yapunov analysis for hybrid dynamical systems.
\newblock In {\em 17th International Conference on Hybrid Systems: Computation
  and Control (part of {CPS} Week), HSCC'14, Berlin, Germany, April 15-17,
  2014}, pages 133--142, 2014.

\bibitem{Kong:13}
H.~Kong, X.~Song, D.~Han, M.~Gu, and J.~Sun.
\newblock A new barrier certificate for safety verification of hybrid systems.
\newblock {\em The Computer Journal}, 57:1033--1045, 2013.

\bibitem{Kozarev:16}
A.~Kozarev, J.~Quindlen, J.~How, and U.~Topcu.
\newblock Case studies in data-driven verification of dynamical systems.
\newblock In {\em Proceedings of the 19th International Conference on Hybrid
  Systems: Computation and Control}, pages 81--86. ACM, 2016.

\bibitem{Kuratko:14}
J.~Ku{\v r}{\'a}tko and S.~Ratschan.
\newblock Combined global and local search for the falsification of hybrid
  systems.
\newblock In A.~Legay and M.~Bozga, editors, {\em Formal Modeling and Analysis
  of Timed Systems}, volume 8711 of {\em Lecture Notes in Computer Science},
  pages 146--160. Springer International Publishing, 2014.

\bibitem{Locatelli:13}
M.~Locatelli and F.~Schoen.
\newblock {\em Global Optimization: Theory, Algorithms, and Applications}.
\newblock SIAM, 2013.

\bibitem{Marti:03}
R.~Mart{\'\i}.
\newblock Multi-start methods.
\newblock In F.~Glover and G.~A. Kochenberger, editors, {\em Handbook of
  Metaheuristics}, volume~57 of {\em International Series in Operations
  Research \& Management Science}, pages 355--368. Springer US, 2003.

\bibitem{Mohri:12}
M.~Mohri, A.~Rostamizadeh, and A.~Talwalkar.
\newblock {\em Foundations of Machine Learning}.
\newblock MIT Press, 2012.

\bibitem{Moore:09}
R.~E. Moore, R.~B. Kearfott, and M.~J. Cloud.
\newblock {\em Introduction to Interval Analysis}.
\newblock SIAM, 2009.

\bibitem{Nieuwenhuis:06}
R.~Nieuwenhuis and A.~Oliveras.
\newblock On {SAT} modulo theories and optimization problems.
\newblock In {\em International conference on theory and applications of
  satisfiability testing}, pages 156--169. Springer, 2006.

\bibitem{Nocedal:99}
J.~Nocedal and S.~J. Wright.
\newblock {\em Numerical optimization}.
\newblock Springer, 2nd edition edition, 2006.

\bibitem{Parrilo:03}
P.~A. Parrilo.
\newblock Semidefinite programming relaxations for semialgebraic problems.
\newblock {\em Mathematical Programming Ser. B}, 96(2):293--320, 2003.

\bibitem{Prajna:04}
S.~Prajna and A.~Jadbabaie.
\newblock Safety verification of hybrid systems using barrier certificates.
\newblock In R.~Alur and G.~J. Pappas, editors, {\em HSCC'04}, number 2993 in
  LNCS. Springer, 2004.

\bibitem{Ratschan:02b}
S.~Ratschan.
\newblock Quantified constraints under perturbations.
\newblock {\em Journal of Symbolic Computation}, 33(4):493--505, 2002.

\bibitem{Ratschan:02f}
S.~Ratschan.
\newblock Efficient solving of quantified inequality constraints over the real
  numbers.
\newblock {\em ACM Transactions on Computational Logic}, 7(4):723--748, 2006.

\bibitem{Ratschan:18}
S.~Ratschan.
\newblock Converse theorems for safety and barrier certificates.
\newblock {\em IEEE Trans. on Automatic Control}, 63(8):2628--2632, 2018.

\bibitem{Sankaranarayanan:08a}
S.~Sankaranarayanan, H.~B. Sipma, and Z.~Manna.
\newblock Constructing invariants for hybrid systems.
\newblock {\em Formal Methods in System Design}, 32(1):25--55, 2008.

\bibitem{Sebastiani:15a}
R.~Sebastiani and S.~Tomasi.
\newblock Optimization modulo theories with linear rational costs.
\newblock {\em ACM Transactions on Computational Logic (TOCL)}, 16(2):12, 2015.

\bibitem{Sebastiani:15}
R.~Sebastiani and P.~Trentin.
\newblock {O}pti{M}ath{SAT}: a tool for optimization modulo theories.
\newblock In {\em International Conference on Computer Aided Verification},
  pages 447--454. Springer, 2015.

\bibitem{Sharma:13}
R.~Sharma, S.~Gupta, B.~Hariharan, A.~Aiken, P.~Liang, and A.~V. Nori.
\newblock A data driven approach for algebraic loop invariants.
\newblock In M.~Felleisen and P.~Gardner, editors, {\em Programming Languages
  and Systems}, volume 7792 of {\em Lecture Notes in Computer Science}, pages
  574--592. Springer Berlin Heidelberg, 2013.

\bibitem{Sloth:12}
C.~Sloth, G.~J. Pappas, and R.~Wisniewski.
\newblock Compositional safety analysis using barrier certificates.
\newblock In {\em HSCC'12}. ACM, 2012.

\bibitem{Sloth:12a}
C.~Sloth, R.~Wisniewski, and G.~Pappas.
\newblock On the existence of compositional barrier certificates.
\newblock In {\em Decision and Control (CDC), 2012 IEEE 51st Annual Conference
  on}, pages 4580--4585, Dec 2012.

\bibitem{SolarLezama:06}
A.~Solar-Lezama, L.~Tancau, R.~Bodik, S.~Seshia, and V.~Saraswat.
\newblock Combinatorial sketching for finite programs.
\newblock {\em SIGPLAN Not.}, 41(11):404--415, 2006.

\bibitem{Taly:09}
A.~Taly and A.~Tiwari.
\newblock Deductive verification of continuous dynamical systems.
\newblock In R.~Kannan and K.~N. Kumar, editors, {\em IARCS Annual Conf. on
  Foundations of Software Technology and Theoretical Computer Science (FSTTCS
  2009)}, volume~4 of {\em Leibniz International Proceedings in Informatics
  (LIPIcs)}, pages 383--394, Dagstuhl, Germany, 2009.

\bibitem{Tarski:51}
A.~Tarski.
\newblock {\em A Decision Method for Elementary Algebra and Geometry}.
\newblock Univ. of California Press, Berkeley, 1951.
\newblock Also in~\cite{Caviness:98}.

\bibitem{Vanecek:96}
A.~Van{\v e}{\v c}ek and S.~{\v C}elikovsk{\'y}.
\newblock {\em Control systems: from linear analysis to synthesis of chaos}.
\newblock Prentice Hall, 1996.

\bibitem{Yang:17}
J.~Yang, M.~A. Islam, A.~Murthy, S.~A. Smolka, and S.~D. Stoller.
\newblock A simplex architecture for hybrid systems using barrier certificates.
\newblock In {\em International Conference on Computer Safety, Reliability, and
  Security}, pages 117--131. Springer, 2017.

\bibitem{Yang:16}
Z.~Yang, C.~Huang, X.~Chen, W.~Lin, and Z.~Liu.
\newblock A linear programming relaxation based approach for generating barrier
  certificates of hybrid systems.
\newblock In J.~Fitzgerald, C.~Heitmeyer, S.~Gnesi, and A.~Philippou, editors,
  {\em FM 2016: Formal Methods: 21st International Symposium}, pages 721--738.
  Springer International Publishing, 2016.

\bibitem{Yang:15}
Z.~Yang, W.~Lin, and M.~Wu.
\newblock Exact safety verification of hybrid systems based on bilinear {SOS}
  representation.
\newblock {\em ACM Trans. Embed. Comput. Syst.}, 14(1):16:1--16:19, Jan. 2015.

\bibitem{Zhou:02}
W.~Zhou, L.~Zhang, and L.~Jiao.
\newblock Linear programming support vector machines.
\newblock {\em Pattern recognition}, 35(12):2927--2936, 2002.

\end{thebibliography}

\Comment{
\appendix

\section{Gradient of Objective Function}

\[ \frac{\partial x^{1/2}}{\partial x}= \frac{1}{2} x^{-1/2}= \frac{1}{2\sqrt{x}}\]

\[ \frac{\partial f(x)^T f(x)}{\partial x_i}= 2 f(x)^T \frac{\partial f(x)}{\partial x_i}\]

\[\frac{\partial ||f(x)||}{\partial x_i}= \frac{\partial (f(x)^T f(x))^{1/2}}{\partial x_i}= \frac{1}{2\sqrt{f(x)^T f(x)}} \frac{\partial f(x)^T f(x)}{\partial x_i}=  \frac{f(x)}{\sqrt{f(x)^T f(x)}} \frac{\partial f(x)}{\partial x_i}= \frac{f(x)}{||f(x)||} \frac{\partial f(x)}{\partial x_i}\] 

\[ \frac{\partial f(x)^T g(x)}{\partial x_i}= \frac{\partial f(x)}{\partial x_i} g(x) + \frac{\partial g(x)}{\partial x_i} f(x) \]

\[ \frac{\partial \frac{f(x)}{||f(x)||}\frac{g(x)}{||g(x)||}}{\partial x_i}= \frac{\partial \frac{f(x)}{||f(x)||}}{\partial x_i}\frac{g(x)}{||g(x)||} + \frac{\partial \frac{g(x)}{||g(x)||}}{\partial x_i}\frac{f(x)}{||f(x)||}\]

\[ \frac{\partial \frac{f(x)}{||f(x)||}}{\partial x_i}= \frac{\frac{\partial f(x)}{\partial x_i}||f(x)||-f(x) \frac{\partial ||f(x)||}{\partial x_i}}{||f(x)||^2}=\frac{\frac{\partial f(x)}{\partial x_i}}{||f(x)||}-\frac{f(x) \frac{\partial ||f(x)||}{\partial x_i}}{||f(x)||^2}= 
\frac{\frac{\partial f(x)}{\partial x_i}}{||f(x)||}-
\frac{f(x) \frac{f(x)}{||f(x)||} \frac{\partial f(x)}{\partial x_i}}{||f(x)||^2}= \]
\[=\frac{\frac{\partial f(x)}{\partial x_i}}{||f(x)||}-
\frac{f(x)^2 \frac{\partial f(x)}{\partial x_i}}{||f(x)||^3}\]
}

\end{document}